\documentclass[journal]{IEEEtran}

\ifCLASSINFOpdf

\else

   \usepackage[dvips]{graphicx}
   \usepackage{amssymb}
   \usepackage{xcomment}

   \graphicspath{{../figures/}}

   \DeclareGraphicsExtensions{.eps}
\fi

\usepackage[cmex10]{amsmath}

\newcommand{\bea}{\begin{eqnarray}}
\newcommand{\eea}{\end{eqnarray}}
\newcommand{\vect}[1]{\mathbf{#1}}
\newcommand{\av}{\mathbf{a}}
\newcommand{\rv}{\mathbf{r}}
\newcommand{\Rv}{{\bf R}}

\begin{document}
%
\title{Information theory of massively parallel probe storage channels.}
%
%
%

\author{Oliver Hambrey, Thomas~Parnell~and~Oleg~Zaboronski
\thanks{O. Hambrey is with the Complexity Doctoral Training Centre, University of Warwick, Coventry, CV4 7AL, UK and
Siglead Europe Ltd., International Digital Laboratory, University of Warwick, Coventry, CV4 7AL, UK e-mail: (oliver.hambrey@siglead.com)}
\thanks{T. Parnell is with Siglead Europe Ltd., UK e-mail: (tom.parnell@siglead.com)}
\thanks{O. Zaboronski is with the Warwick Mathematics Institute, University of Warwick and Siglead Europe Ltd. e-mail: (O.V.Zaboronski@warwick.ac.uk)}}
%
%

\markboth{IEEE Transactions on Information Theory,~Vol.XX, No.X , January~2011}%
{Shell \MakeLowercase{\textit{et al.}}: Bare Demo of IEEEtran.cls for Journals}

\maketitle

\begin{abstract}
Motivated by the concept of probe storage,
we study the problem of information retrieval using a large array of $N$ nano-mechanical probes, $N\sim 4000$. At the nanometer 
scale it is impossible to avoid errors in the positioning of the array, thus all signals retrieved by the probes of the array at a 
given sampling moment are affected by the same amount of random position jitter. Therefore a massively parallel probe storage
device is an example of a noisy communication channel with long range correlations between channel outputs due to the global
positioning errors.

We find that these correlations have a profound effect on the channel's properties. For example, it turns out
that the channel's information capacity does approach $1$ bit per probe in the limit of high signal-to-noise ratio, but
the rate of the approach is only polynomial in the channel noise strength. Moreover,  $any$ error
correction code with block size $N>>1$ such that codewords correspond to the instantaneous
outputs of the all probes in the array exhibits an error floor independently of the code rate. We illustrate
this phenomenon explicitly using Reed-Solomon codes the performance of which is easy to simulate numerically.

We also discuss capacity-achieving error correction codes for the global jitter
channel and their complexity.

\end{abstract}

\IEEEpeerreviewmaketitle

\section{Introduction}

\IEEEPARstart{T}{he} invention of atomic force microscopy in $1986$ \cite{Binnig1986} opened the possibility of storing information at nanometer
scales resulting in proposals for achieving aerial information densities of tens or even hundreds of Terabits per
square inch. The basic idea is that information can be stored by altering certain features of the storage
medium at the scale of nanometers. These changes can be then sensed and information retrieved by nanoscale
probes similar to those used in atomic force microscopy. For example, binary information can
be stored in crystalline dots created in amorphous media or amorphous dots in crystalline media
and retrieved using electric probes (phase change storage, \cite{Wright2009}). Alternatively, the information can be stored
using indentations or, more recently, variable length grooves made in polymer media and retrieved using thermoelectric probes
(thermo-mechanical probe storage, \cite{Mamin1998}).

\begin{figure}
\centering
\includegraphics[width=3.5in]{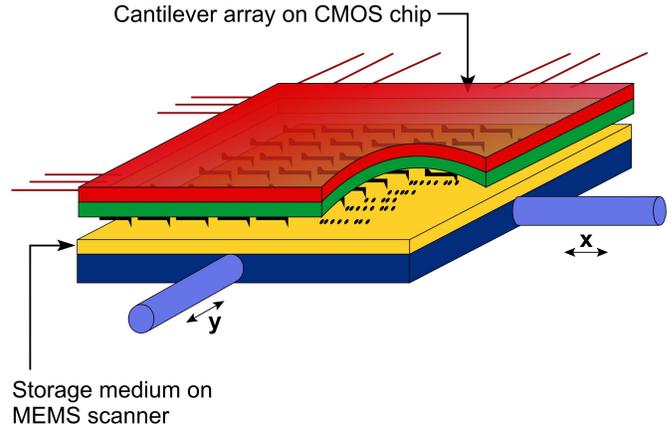}
\caption{The layout of the 'Millipede' - a thermomechanical probe storage device. Courtesy of IBM - Zurich, \cite{Despont1999}}
\label{fig:millipede}
\end{figure}

\begin{figure}
\begin{center}
\includegraphics[width=3.3in]{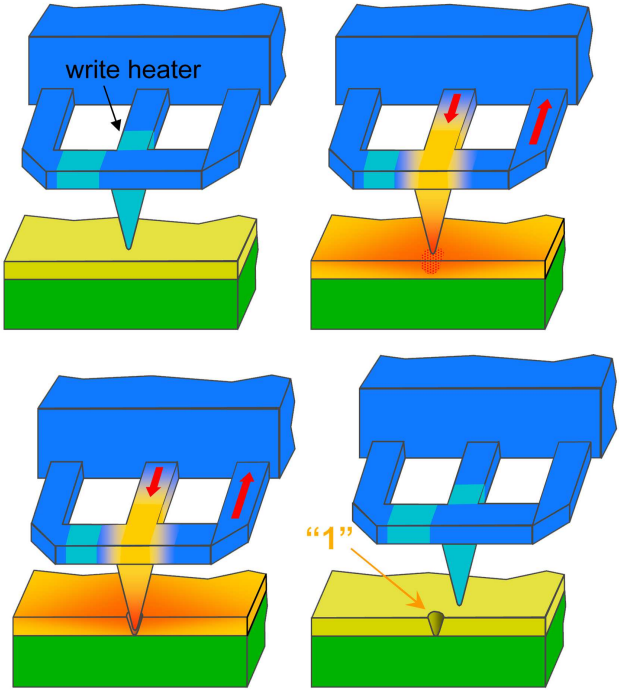}
\caption{The principle of thermomechanical reading and writing. Top left picture, writing bit 'zero':
cold probe pressed against the polymer surface leaves no mark in the media. Top right picture,
writing bit 'one': the probe heated above the polymer's melting temperature and pressed against the polymer surface
leaves an indentation in the media shown in the bottom left picture. Bottom right pictures, reading bits 'one' and 'zero': the probe
inserted in the indentation is cooler than the probe pressed against the flat surface. These temperature variations can be captured using
thermo-resistive sensors. Courtesy of IBM Zurich, \cite{Vettiger2002} }
\label{fig:read_write}
\end{center}
\end{figure}

The whole concept of storing and retrieving information using nano-scale probes became known as probe storage, see \cite{Leon2011}, Chapter 4,
for a comprehensive review of the current state of the field. Perhaps the most widely known concept of the
probe storage device is the IBM's 'Millipede' \cite{Vettiger2002}, \cite{Pozidis2003} - an array of thermo-electric probes with sharp tips used
to create and sense indentations in the polymer media. The layout of the Millipede is shown in Fig. \ref{fig:millipede}, the basic
principle of thermo-mechanical reading and writing is explained in Fig. \ref{fig:read_write}.
It has been demonstrated that data can be retrieved with low bit error rate per probe (about $10^{-4}$) at densities of up to $2$ Terabit per square inch \cite{Pozidis2010}.

The particular feature of the Millipede shared by all existing concepts of probe storage is the presence of
a large ($\sim2^{12}$) number of probes reading and writing the information in parallel. This feature makes probe storage
very different from more traditional storage devices such as magnetic or optical disks or flash memory. Each probe reads and writes
information in its own field. The array of probes moves as a whole to allow each probe to explore its field. For aerial densities of information
of several terabit per square inch,
the array has to be moved by a distance of the order of $10$ nanometers from one set of the sampling points to the next and repositioned with a sub-nanometer precision for writing or reading.
Inevitably, positioning errors affect every single probe in the array. We refer to the combined effect of errors in the positioning of the array at both the reading and writing stages
as global positioning jitter or just global jitter.
The aim of this paper is to investigate the performance of error correcting codes operating on the output of the probe array subject to highly correlated disturbances due to global
jitter. We will investigate both the performance of special codes (Reed-Solomon) and address the general question of existence of good error correction codes for the global
jitter channel by calculating channel capacity and studying  Gallager's random coding bound.

Despite the fairly abstract tools used in the paper, the purpose of our investigation is to answer a very practical question: given a 'good' communication channel
(e. g. the single probe's channel with bit error rate $10^{-4}$), how good is a communication system (e.g. a Millipede) consisting of thousands of good channels subjected
to correlated error events?

Our system-level analysis allows one to understand advantages and limitations of complicated communications devices without actually
building one. In particular, it turns out that information theoretical performance limits derived in this paper have crucial system-level implications for
the design of error correction codes for probe storage.

The rest of the paper is organised as follows. In Section \ref{sec:chmod} we introduce a simple model for the probe storage
channel, the global jitter channel,  which accounts for the effects of the probe array's position jitter and calculate the probability distribution
function of signal amplitude for Gaussian isolated pulse response and Gaussian statistics of jitter. In Section \ref{sec:opdet} we introduce a low
complexity signal detection scheme for the global jitter channel
and prove its optimality in the limit of large array sizes.
In Section \ref{sec:rs} we calculate block error rate for non-interleaved Reed-Solomon
codes applied to the probe storage channel and analyze their error floor behaviour. In Section \ref{sec:cap} we calculate Shannon's capacity of the
global jitter channel and show that it approaches $1$ bit according to a power law
in the limit of large signal-to-noise ratio. In Section \ref{sec:rcb} we calculate the average block error rate
for non-interleaved codes sampled from Gallager's ensemble (the random coding bound or RCB) and show that it
exhibits an error floor behaviour as a function of SNR. In Section \ref{sec:fano} we show that there exists
no non-interleaved codes with a positive rate which can be used for error-free retrieval of information using large arrays of probes in the
presence of global jitter. In Section \ref{sec:goodcodes} we discuss minimal requirements for capacity-achieving
codes for global jitter channels. We conclude our work with Section \ref{sec:concl} which contains the summary
of the results of our investigation. For completeness, we present the derivations of more technical results obtained
in the paper in the Appendix.

\section{Channel Model}\label{sec:chmod}
We consider the system consisting of an array of $N$ probes reading/writing in parallel. We assume that channel coding has been
used (an RLL code for example) and that the symbol pitch is large enough so that inter-symbol interference can be ignored.
The sampled readback signal at the $k$-th probe at the $t$-th moment in time is modelled as:
\begin{eqnarray}
r^{(k)}_t = p(J_t)a^{(k)}_t + \sigma n^{(k)}_t, ~k=1,2,\ldots, N\label{chan:glob_jit}
\end{eqnarray}
where $a^{(k)}_t\in\{0,1\}$ is the bit written to the medium by the $k$-th probe at time $t$, $J_t$ is the global
positioning error (jitter) at time $t$, $p(J)$ is the channel impulse response and $\{\sigma n^{(t)}_k\}$ is a sequence of
random variables modelling the combined effect of electronics and media noise.

Experiments carried out at IBM Zurich \cite{Pozidis2007} have confirmed that for thermo-mechanical storage media the combined electronics/media
noise is well modelled by Gaussian random variables. Thus we assume that $\{n^{(k)}_t\}$ are independent
identically distributed Gaussian variables with mean zero and unit variance.
Parameter $\sigma$ is the standard deviation of the resulting additive white Gaussian noise (AWGN).

It has also been demonstrated experimentally that when using a Millipede-like positioning
control loop the random jitter $J_t$ is also well modeled by a mean-zero Gaussian random variable \cite{Pozidis2007}.
Note that this observation applies to any probe storage device, for instance a system based on phase
change media, as long as a similar positioning system is used. Let $\sigma_J$ be the standard deviation of Gaussian jitter.

Most of the qualitative results reported below do not depend on the detailed assumptions about the statistics of jitter or the precise shape of the impulse
response. For quantitative analysis and numerical simulations of the global jitter channel
we will use the Gaussian impulse response:
\begin{eqnarray}
p(J) = e^{-\frac{J^2}{W^2}}\label{eq:ex_imp}
\end{eqnarray}
where $W$ is a parameter related to pulse width.
In what follows it will be more convenient to work directly with the random variable $p(J)\in (0,1)$ which measures signal
amplitude degradation due to position jitter $J$. A calculation leads to the following answer for the probability density function of $p$:
\begin{eqnarray}
\rho_P(p)\equiv \mathbf{E}_J \delta(p-p(J))=\frac{W}{\sqrt{2\pi \sigma_J^2\log\left(\frac{1}{p}\right)}}p^{\frac{W^2}{2\sigma_J^2}-1}.
\label{eq:pdf}
\end{eqnarray}
Finally we define the signal-to-noise ratio (SNR) of the global jitter channel measured in decibels as follows:
\begin{eqnarray}
SNR = 10\log_{10}\left(\frac{1}{\sigma^2}\right)
\end{eqnarray}

\section{Optimal Channel Detector}\label{sec:opdet}

The global jitter channel is characterised by strong equal time correlations between outputs of all $N$-probes in the array.
In the limit $N\rightarrow\infty$ it is
possible to exploit these correlations and derive an asymptotically optimal low-complexity detection scheme.

To motivate the rigorous argument given below, let us ask the following question: what is the optimal channel detector conditional on the knowledge of
the value of jitter at time $t$? Conditional on the known amplitude value $p_t\equiv p(J_t)=p$, channel outputs $r_t^{(k)}$'s are independent. Consequently,
the optimal maximum a posteriori (MAP) detector \cite{Kay1998} is simply the collection of $N$ independent optimal threshold
detectors for AWGN channel:
\begin{eqnarray}
\hat{a}_t^{(k)} = \left\{\begin{array}{cc}
1 & \mbox{if $r_t^{(k)} > \frac{p}{2}$ } \\
0 & \mbox{if $r_t^{(k)} < \frac{p}{2}$}\
\end{array}\right.
\label{eq:genie}
\end{eqnarray}

The problem with the above 'Genie-assisted' detector is that the value of
the amplitude $p$ is a priori  unknown. However the value of $p$ can be reliably estimated from
the string $r^{(1)}_t, r^{(2)}_t, \ldots, r^{(N)}_t$ if $N$ is sufficiently large.
The key
observation concerns the sample average of all $N$ received signals at time $t$:
\begin{eqnarray}
\overline{R}_t &=& \frac{1}{N}\sum_{k=1}^Nr_t^{(k)}\nonumber\\
&=& p(J_t)\sum_{k=1}^N\frac{a_t^{(k)}}{N} + \sigma\sum_{k=1}^N\frac{n_t^{(k)}}{N}\label{eq:sum_of_samp_avs}
\end{eqnarray}
Assume that channel inputs $a_t^{(k)}$ are independent and uniformly distributed. Then by the strong law of large numbers \cite{Durret2004}, both sample
averages in (\ref{eq:sum_of_samp_avs}) converge almost surely to their expected values as $N\rightarrow\infty$:
\begin{align}
\sum_{k=1}^N\frac{a_t^{(k)}}{N}& \overset{a.s.}{\longrightarrow}\frac{1}{2}  & \sum_{k=1}^N\frac{n_t^{(k)}}{N}& \overset{a.s.}{\longrightarrow} 0
\end{align}
Substituting into (\ref{eq:sum_of_samp_avs}) we find in the large $N$ limit it is actually possible to compute the previously unknown distortion $p_t=p(J_t)$ as follows:
\begin{eqnarray}
p_t = \lim_{N\rightarrow\infty}\left(2\overline{R}_t\right)\label{eq:sig_deg_due_to_jit}
\end{eqnarray}
Feeding this estimate of $p$ to $N$ independent threshold detectors (\ref{eq:genie}) we obtain estimates $\hat{a}_1, \hat{a}_2, \ldots, \hat{a}_N$ of data bits which should become
optimal in the limit $N\rightarrow \infty$.

The heuristic argument given above leads to the following asymptotically optimal detection algorithm for the global jitter channel:

\begin{enumerate}
\item {\bf Input:} Channel output $r^{(1)}_t, r^{(2)}_t, \ldots r^{(N)}_t$ at time $t$.
\item Estimate the threshold:
\bea
T_N=\frac{1}{N} \sum_{k=1}^N r^{(k)}_t\label{eq:thr}
\eea
\item Perform bit-by-bit detection:
\begin{eqnarray}
\hat{\hat{a}}_t^{(k)} = \left\{\begin{array}{cc}
1 & \mbox{if $r_t^{(k)} > T_N$ } \\
0 & \mbox{if $r_t^{(k)} < T_N$ }\
\end{array}\right.
\label{eq:det}
\end{eqnarray}
\item {\bf Output:} The estimate of channel input $\hat{\hat{a}}^{(1)}_t \hat{\hat{a}}^{(2)}_t \ldots \hat{\hat{a}}^{(N)}_t$.
\end{enumerate}

We will now prove that detector (\ref{eq:det}) is indeed optimal in the limit $N\rightarrow \infty$.
Firstly, note that detector (\ref{eq:genie}) is an optimal MAP detector which infers the most probable channel input conditional on ${\bf{r}_{t},p_t}$. Therefore
its bit error rate is smaller or equal than bit error of any detector which infers channel input conditional on channel input $\bf{r}_t$ only.
So we can establish the asymptotic optimality of (\ref{eq:det}) by proving that its bit error rate approaches the bit error rate of
(\ref{eq:genie}) in the limit $N\rightarrow \infty$:
\bea
\lim_{N\rightarrow \infty} |\Pr(\hat{\hat{a}}_t^{(k)}\neq a_t^{(k)})-\Pr(\hat{a}_t^{(k)}\neq a_t^{(k)})|=0.
\label{eq:claim}
\eea
A long but straightforward calculation presented in Appendix \ref{ap:bound} shows that
\bea
|\Pr(\hat{\hat{a}}_t^{(k)}\neq a_t^{(k)})-\Pr(\hat{a}_t^{(k)}\neq a_t^{(k)})|\leq \frac{3\left(1 + \frac{\mathbb{E}(p^2)}{4\sigma^2}\right)^{1/3}}{(2\pi N)^{1/3}}.
\label{eq:bound}
\eea
Taking the large-$N$ limit of both sides of the above inequality we arrive at (\ref{eq:claim}).
Thus the asymptotic optimality of (\ref{eq:det}) is established. We will refer to detector (\ref{eq:det}) as the {\em $LLN$-detector}
and detector (\ref{eq:genie}) as $Genie-detector$.

\begin{figure}
\begin{center}
\includegraphics[width=3.5in]{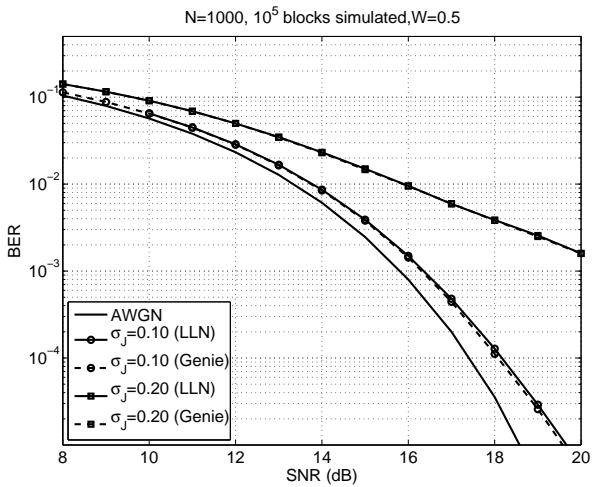}
\caption{BER performance comparison of LLN detector against the ideal detector with perfect knowledge of global jitter}
\label{fig:ber_lln}
\end{center}
\end{figure}

The performance of the LLN-detector for large but finite $N$ is compared with the performance of Genie-detector in
Fig. \ref{fig:ber_lln} for Gaussian pulse (\ref{eq:ex_imp}) and various values of jitter strength $\sigma_J$. The size of the
probe array is $N=1000$, the pulse width is $W=0.5$ and the simulation has been run for $10^{5}$ consecutive
readings of the entire probe array. As a reference the performance of the channel with no jitter which
corresponds to the classical binary-input additive white Gaussian noise (AWGN) channel has been included.
It is observed that as jitter strength is increased the BER performance degrades significantly but for all
cases the performance of the LLN-detector is virtually indistinguishable from Genie-assisted detector. It is interesting
to note that the gap between BER of Genie-detector and the LLN-detector increases with SNR, see Fig. \ref{fig:ber_lln}.
This observation is consistent with the bound (\ref{eq:bound}): The right hand side of the bound grows if the block
size $N$ is kept fixed but the additive noise strength $\sigma$ is reduced. The intuition behind the observed divergence
is very simple: for smaller noise one needs a more precise threshold estimate to stay
near the performance of the Genie-detector. The precision of the threshold estimate depends
on the rate of convergence of the sample sum and scales as $1/\sqrt{N}$ in accordance with Central Limit Theorem (see \cite{Durret2004} for a review).

It is interesting to compare the LLN detector to the simplest detection scheme proposed for Millipede, see. \cite{Pozidis2005}. The Millipede
detector consists of $N$ independent MAP detectors (one detector per probe). The $k$-th detector estimates
the data bit $a^{(k)}_t$ at time $t$ conditional on the output $r_t^{(k)}$ only. Unlike the LLN-detector, there is no sharing of information between individual
detectors.

The maximum-likelihood detector of the single probe output modeled by (\ref{chan:glob_jit}) is a
simple threshold detector: the most likely bit given the  output of the $k$-th probe, $\hat{a}_t^{(k)}$ is given by:
\begin{eqnarray}
\hat{a}_t^{(k)} = \left\{\begin{array}{cc}
1 & \mbox{if $r_t^{(k)} > r_0$ } \\
0 & \mbox{otherwise }\
\end{array}\right.
\end{eqnarray}
where $r_0$ is the optimal threshold - a number between $0$ and $1$  which can be either measured experimentally or computed theoretically
for a given the channel model by solving the maximum likelihood equation
$$
\Pr(r_t^{(k)}=r_0\mid a_t^{(k)}=1)=\Pr(r_t^{(k)}=r_0\mid a_t^{(k)}=0),
$$
see \cite{Parnell2009} for more details.

\begin{figure}
\begin{center}
\includegraphics[width=3.5in]{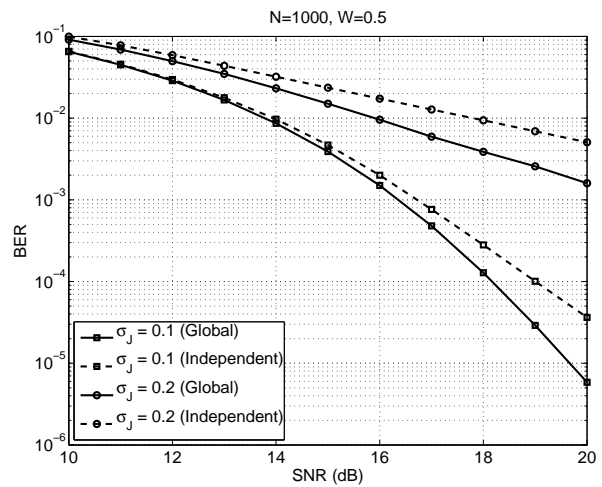}
\caption{Comparison of BER performance of the LLN-detectors for the global jitter channel and the set of $N$-independent threshold detectors}
\label{fig:ber_compare}
\end{center}
\end{figure}

In Figure \ref{fig:ber_compare} the BER of the Millipede detection scheme is compared with that of the LLN-detector
for two values of jitter strength $\sigma_J$. The result is quite striking: the optimal
detector for the global jitter channel outperforms the set of $N$ independent threshold detector by over a decibel for BER$\sim 10^{-4}$.

Finally, let us analyse the complexity of the LLN-detector. The complexity of
adding $N$  fixed precision numbers in (\ref{eq:thr}) scales as $Nlog(N)$, the complexity of the detection step (\ref{eq:det}) is $O(N)$ so the
overall detection complexity is $O(Nlog(N))$, the detection complexity per detected bit is $O(log(N))$.

Having constructed the optimal detection scheme we can investigate the performance of error correction codes for the global jitter
channel starting with Reed-Solomon codes which featured in the original Millipede proposal.

\section{The performance of Reed-Solomon Codes for the global jitter channel.}\label{sec:rs}

The typical probe storage array size $N$ is $64\times 64=4096$, which is close to the sector size of the previous generation of hard disk drives.\footnote{January 2011 has been
designated as the date of the final transition from 512B sector size in HDD's to the 4KB sector size.}  The simplest error correction coding (ECC) scheme
for probe storage follows the example of hard drives: for the latter data is encoded sector-by-sector, for the former ECC is applied independently
to $K$-bit strings of data $\mathbf{I}_1,\mathbf{I}_2,\mathbf{I}_3\ldots$
to be recorded on the media by the probe array at times $t_1, t_2, t_3, \ldots$. We refer to the described application of error correction
coding as {\em non-interleaved} meaning that channel outputs corresponding to different moments of sampling time cannot belong to the same
code block.

The non-interleaved ECC block size is equal to the number of probes $N$, the ECC rate is $R=K/N$.
During the reading stage, the single-time output $\hat{a}_t^{(1)}, \hat{a}_t^{(2)},\ldots , \hat{a}_t^{(N)}$ of the channel detector is fed into the ECC decoder resulting
in an estimate $\hat{\mathbf{I}}_t$ of $K$ recorded bits at each sampling time $t$.

We start our study of error correction for global jitter channel with classical Reed-Solomon (RS) codes, see \cite{Lin2004} for review.
These are the $(N_s,K_s)$ symbol block codes over the Galois field $GF(2^n)$. Here $N_s=N/n$ is the number of RS symbols per block, $K_s=K/n$
is the number of information symbols. Symbols are represented by $n$-bit binary strings. The maximal block size of the code is $N_{max}=2^n-1$ symbols. An RS code
with any block size $N_s\leq N_{max}$ can be constructed by treating missing $(N_{max}-N_s)$ symbols as zeros (this operation is called shortening).
A Reed-Solomon code with rate $R=K_s/N_s$ can correct up to $$N_s\frac{1-R}{2}$$ incorrectly detected symbols, which makes it a maximal distance separable (MDS) code.

The fact that the error event of a RS code depends on the number of incorrectly detected symbols rather than bits, makes it a very good code for
channels dominated by relatively short bursts of noise. It is therefore not very well suited for the global jitter channel. However we will see below that conclusions
drawn from analyzing the performance of RS codes in the presence of global jitter can be applied to $any$ non-interleaved block ECC.

\begin{figure}
\begin{center}
\includegraphics[width=3.5in]{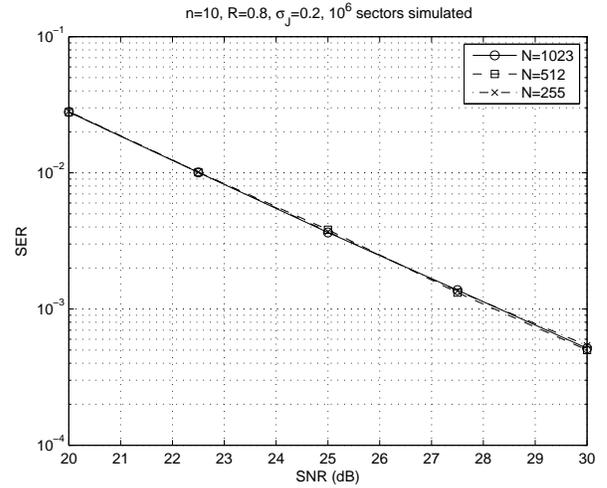}
\end{center}
\caption{RS SER curves do not depend on the block size and exhibit an error floor behaviour.}
\label{fig:rs_blocksize}
\end{figure}

We start with presenting results of numerical simulations.
Fig. \ref{fig:rs_blocksize} illustrates the performance of rate-$0.8$ RS codes with symbol size $n=10$ bits. Three codes
are considered: with block size $N_s=1023$ (the maximal block size) and two shortened RS codes - with $N_s=511$ and  $N_s=255$.
These block sizes correspond to hypothetical probe arrays with $2550$, $5110$ and $10230$ probes.
To model global jitter we use the channel (\ref{chan:glob_jit}) with jitter strength $\sigma_j=0.2$, and pulse width $W=0.5$.
For each of the codes we measure the probability of sector error or sector error rate (SER) as a function of signal-to-noise ratio (SNR).
SER is measured as the number of times the number of symbol errors in the received string exceeded RS threshold $N_s\frac{1-R}{2}$.
The total number of strings is $10^6$ to ensure that the total error count used to estimate SER is at least $10^2$ for the highest SNR point.

The first striking feature of the curves presented in Fig. \ref{fig:rs_blocksize} is that there
no discernible performance loss using the
shortened codes. This is in stark contrast from the known behaviour of RS codes for the AWGN channel where
$$
\frac{logSER_{RS(N_1,R)}}{logSER_{RS(N_2,R)}}\stackrel{N\rightarrow \infty} {\longrightarrow}\frac{N_1}{N_2}.
$$

Therefore, the numerical evidence suggests that the probability of RS codes applied to the global jitter channel does $not$ depend on the code's block size.

The second unusual feature of SER vs. SNR curves shown in Fig. \ref{fig:rs_blocksize} is the exponential law of the decay of SER with SNR - all the curves
look like straight lines on the semi-logarithmic plot. The exponential rather than the 'waterfall' shape of SER curves is often referred to as an 'error floor'.
Its appearance is normally attributed to the use of suboptimal decoding algorithms such as belief propagation, rather than channel properties.  Here we are driven to
a conclusion that non-interleaved RS codes decoded with an optimal maximum likelihood hard input algorithm exhibit an error floor, the position of which is independent of the block size.

As it turns out, the breakdown of Reed-Solomon codes for channels with global jitter
can be understood analytically using the machinery of large deviations, \cite{Dembo1993}.

To achieve this, we need to introduce some notations.
Let \bea\tau=(1-R)/2\label{eq:rst}\eea be the maximal fraction of correctable symbols for the code at hand.
Let the random variable $\xi_k$ take values $\{0,1\}$ with probabilities $\{e_0,e_1\}$ respectively. Here $e_1$ is the probability that the $k$-th
symbol is detected incorrectly. Due to statistical homogeneity of the channel model (\ref{chan:glob_jit}) the probability of symbol error does not depend
on the symbol index $k$. The event $\xi_k=1$ corresponds to the
$k$-th symbol being detected erroneously and $\xi_k=0$ the $k$-th symbol being detected correctly.
Conditionally on the event $p_t=p$ the probability of RS sector error is simply given by a multinomial formula \cite{Evans2000}. Unfortunately, the multinomial
formula is not very useful for quantitative analysis in the region of low SER. Instead, we are going to use simple asymptotic expressions for RS sector error rate based on Cramer's
theory \cite{Dembo1993}.
The event of sector error corresponds to the fraction of incorrectly detected symbols exceeding RS threshold (\ref{eq:rst}).
Hence, the probability of a block being decoded incorrectly can be written:
\begin{eqnarray}
\Pr(SE\mid p_t=p) = \Pr\left(\sum_{k=1}^{N_s}\xi_k > \tau N_s \mid p_t=p\right)
\end{eqnarray}
An application of Chernoff's bound \cite{Durret2004} results in the following upper bound on $\Pr(SE)$ for any $\lambda > 0$:
\begin{eqnarray}
\Pr(SE\mid p_t=p) \leq e^{-\lambda\tau N_s}\mathbb{E}\left[e^{\lambda\left(\sum_{k=1}^{N_s}\xi_k\right)}\mid p_t=p \right],\label{eq:chernoff}
\end{eqnarray}
where $\mathbb{E} [\bullet \mid p_t=p ]$ stands for $p_t$-conditional expectation value.
Recall that conditional on the value of signal amplitude $p_t$ the global jitter channel is memoryless. Therefore symbol error events $\xi_k$ are
conditionally independent
and identically distributed and the bound in (\ref{eq:chernoff}) can be re-written as:
\begin{eqnarray}
\Pr(SE\;|\;p_t = p) \leq e^{-\lambda\tau N_s + N_s\ln\mathbb{E}\left[e^{\lambda \xi}\mid p_t=p\right]}\label{eq:chernoff_cond}
\end{eqnarray}
The bound in (\ref{eq:chernoff_cond}) holds for any $\lambda > 0$ and thus we can choose the tightest bound possible by minimising over
all $\lambda$'s:
\begin{eqnarray}
\frac{1}{N_s}\ln\Pr(SE\;|\;p_t=p) \leq I(\tau, p) \label{eq:chernoff_rate}
\end{eqnarray}
 Function $I(\tau,p)$  is known in the theory of large deviations as the rate function and is given by:
\begin{eqnarray}
I(\tau,p) = \inf_{\lambda > 0}\left( -\lambda \tau + \ln\mathbb{E}\left[e^{\lambda\xi}\mid p_t=p \right]\right) \label{eq:rate_function}
\end{eqnarray}
The question remains: is the bound in (\ref{eq:chernoff_rate}) tight enough to be useful? The answer is provided by an
application of Cramer's Theorem \cite{Dembo1993} which states that provided $\mathbb{E}\left[\xi\mid p_t=p \right] < \tau$ (which is certainly true in
the limit of low sector error-rates) the bound (\ref{eq:chernoff_rate}) is tight in the limit of large block size:
\begin{eqnarray}
\lim_{N_s\rightarrow\infty}\frac{1}{N_s}\ln\Pr(SE\;|\;p_t=p) = I(\tau, p)
\end{eqnarray}
An explicit expression for the rate function (\ref{eq:rate_function}) can be found by solving the critical point equation:
\begin{eqnarray}
\frac{\partial}{\partial \lambda}\left(-\lambda \tau + \ln\mathbb{E}\left[e^{\lambda\xi}\mid p_t=p \right]\right) = 0\label{eq:cpe}
\end{eqnarray}
The function differentiated in (\ref{eq:cpe}) is convex with the unique point of global minimum given by:
\begin{eqnarray}
\lambda_c = \ln\left(\frac{e_0(p)}{e_1(p)}\frac{\tau}{(1-\tau)}\right),
\end{eqnarray}
where $e_1(p)=\mathbb{E}(\xi\mid p_t=p)$ is the conditional symbol error rate, $e_0(p)=1-e_1(p)$.
If the condition of Cramer's theorem $\mathbb{E}\left[\xi\mid p_t=p \right] = e_1(p) < \tau$ is satisfied then $\lambda_c > 0$ and
substituting back into (\ref{eq:rate_function}) we find the rate function can be expressed as a Kullback-Leibler divergence \cite{David2003}:
\begin{eqnarray}
I(\tau,p) = -D_{KL}\left(\left(1-\tau,\tau\right) || \left(1-e_1(p),e_1(p)\right)\right)
\end{eqnarray}
Recall that for any two stochastic vectors $\mathbf{P}$ and $\mathbf{Q}$,
\begin{eqnarray}
D_{KL}(\mathbf{P} || \mathbf{Q})=\sum_{k}P_k \log\left( \frac{P_k}{Q_k} \right)
\end{eqnarray}
On the other hand if $e_1(p) \geq \tau$ then $\lambda_c \leq 0$ and due to convexity the minimum of (\ref{eq:rate_function})
is achieved at $\lambda=0$ resulting in the trivial bound $I(\tau,p)=0$.

Let us summarise our findings so far:
\begin{eqnarray}
\Pr(SE\mid p_t=p) \lesssim e^{-N_sI(\tau,p)},
\label{eq:rf_cond1}
\end{eqnarray}
where
\begin{equation}
I(\tau,p)=
\left\{ \begin{array}{ll}  D_{KL}\left(\left(1-\tau,\tau\right) || \left(e_0(p),e_1(p)\right)\right) &\mbox{if } e_1(p)<\tau\\
0&\mbox{if } e_1(p)>\tau
\end{array} \right.
\label{eq:rf_cond2}
\end{equation}

Now we can derive an upper-bound for the performance of non-interleaved RS codes in the global jitter channel
using the upper bound (\ref{eq:rf_cond1}, \ref{eq:rf_cond2}) on the conditional SER:
\begin{eqnarray}
\Pr\left(\;SE\;\right) &=& \int_0^1 d\mu_P(p)\; \Pr\left(\;SE\;|\;p_t=p\right),\label{eq:p_se}
\end{eqnarray}
where $d\mu_P(p)=\rho_P(p)dp$ is the probability measure of jitter-dependent signal amplitude with density (\ref{eq:pdf}). The range of integration can now be
split around the critical value of signal degradation $p_c$ which is the unique solution of the equation
$$ e_1(p_c)=\tau.$$
Then using the upper bound (\ref{eq:rf_cond1}) for the conditional probability of RS error we find:
\begin{eqnarray}
\Pr\left(\;SE\;\right) \leq \int_0^{p_c}d\mu_P(p) + \int_{p_c}^1\;d\mu_P(p)\;e^{-N_sF_\tau(p)}\label{eq:split_int}
\end{eqnarray}
Where $F_\tau(p)$ is expressed in terms of Kullback-Leibler divergence:
\begin{eqnarray}
F_{\tau}(p) = \operatorname{D_{KL}}\left(\left(1-\tau,\tau\right) || \left(e_0(p),e_1(p)\right)\right)\nonumber
\end{eqnarray}
Note that $F_\tau(p)$ has a unique non-degenerate minimum at $p_c$ where it takes the value $F(p_c)=0$. The proof of this
fact follows easily from Gibbs inequality \cite{David2003}.
Therefore, we can apply the Laplace formula \cite{Erdelyi1956} to the second integral in equation (\ref{eq:split_int}) to derive the large-$N_s$ asymptotic of the probability
of sector error:
\begin{eqnarray}
\int_{p_c}^1\;d\mu_P(p)\;e^{-N_s\;F(p)} \sim  \rho_P(p_c)\sqrt{\frac{\pi}{N_sF''(p_c)}}\label{eq:laplace_result}
\end{eqnarray}
Note that the resulting expression (\ref{eq:laplace_result}) tends to zero as $\sqrt{\frac{1}{N_s}}$ in the limit $N_s\rightarrow\infty$.
Therefore what we have discovered is that for the global jitter channel in the limit $N_s\rightarrow\infty$  the probability of sector error is upper-bounded as follows:
\begin{eqnarray}
\Pr\left(\;SE\;\right) \lesssim \Pr( p_t \leq p_c) + \rho_P(p_c)\sqrt{\frac{\pi}{N_sF''(p_c)}}\label{eq:rsup}
\end{eqnarray}
We therefore confirmed theoretically that for a fixed level of noise and in the limit of large sector size, the probability of
sector error for information encoded with a Reed-Solomon code and transmitted over the global jitter channel does not depend on the code's block size.
This conclusion is in perfect agreement with the results of numerical simulations shown in Fig. \ref{fig:rs_blocksize}.

Using (\ref{eq:rsup}) we can also explain the shape of the SER vs SNR curves thus confirming the appearance of the error floor analytically.
Moreover, we will be able to determine the position of the error floor as a function of code rate  $R=1-2\tau$.

Up to this point our considerations did not depend on the specific shape of the impulse response or on jitter statistics. From now on we will assume
the Gaussian impulse response given by equation (\ref{eq:ex_imp}) and Gaussian position jitter. Then it follows from (\ref{eq:rsup})
that
\begin{eqnarray}
\Pr(SE) \lesssim \Pr(p_t\leq p_c)+O(N_s^{-1/2}) \nonumber\\= \frac{2}{\sqrt{2\pi\sigma_J^2}} \int_{J_c}^{\infty}dJ\; e^{-\frac{J^2}{2\sigma_J^2}}+O(N_s^{-1/2})\label{eq:erf}
\end{eqnarray}
where $J_c = W\sqrt{\ln(p_c^{-1})}$ is the critical value of jitter that causes signal degradation $p_c$.
Bounding the integral in equation (\ref{eq:erf}) with elementary functions we arrive at the following:
\begin{eqnarray}
\Pr(SE) \leq p_c^{\left(\frac{W^2}{2\sigma_J^2}\right)}\label{eq:rsup2}
\end{eqnarray}
If we further assume that a high-rate Reed-Solomon is used so that $\tau \ll 1$ then it follows that the symbol error rate
conditional on $p_c$, $e_1(p_c)=\tau$ is also much less that $1$. In this limit we can approximate symbol error rate as follows:
\begin{eqnarray}
e_1(p_c) = 1 - (1-f(p_c))^n \approx n f(p_c)
\end{eqnarray}
where $f(p_c)$ is the bit error rate conditional on $p_c$ which admits the following upper bound:
\begin{eqnarray}
f(p_c) &\leq& \frac{1}{2}\; e^{-\frac{p_c^2}{8\sigma^2}}
\end{eqnarray}
Note that this upper bound is tight in the limit of large signal to noise ratios - the region relevant for studying the error floor.
Therefore we can estimate the critical point $p_c$ by $\hat{p}_c \geq p_c$ as follows:
\begin{eqnarray}
\hat{p}_c = \sigma \sqrt{8\ln\left(\frac{n}{2\tau}\right)}\label{eq:pc_for_rs}
\end{eqnarray}
Substituting (\ref{eq:pc_for_rs}) into equation (\ref{eq:rsup2}) we arrive as the following bound on sector error rate for the global jitter
channel with the exponential impulse response for high-rate Reed-Solomon codes:
\begin{eqnarray}
\Pr(SE) &\lesssim& \left(8\ln\left[\frac{n}{2\tau}\right]\right)^{\left(\frac{W^2}{4\sigma_J^2}\right)}\;\sigma^{\left(\frac{W^2}{2\sigma_J^2}\right)}+O(N^{-1/2}).\nonumber\\
\label{eq:rs_ef_th}
\end{eqnarray}
This is a disastrous result: the probability of sector error decays as power law in $\sigma$ with an exponent that does not
depend on code rate or block size! The algebraic dependence of SER on the noise strength implies the exponential
dependence of the probability of sector error on signal-to-noise ratio which explains straight lines on the semi-logarithmic SER-SNR plot in Fig. \ref{fig:rs_blocksize}.

Moreover, we observe that the position of the error floor (determined by the pre-factor in (\ref{eq:rs_ef_th})) depends very weakly (logarithmically) on
the code rate $R=1-2\tau$.

\begin{figure}
\begin{center}
\includegraphics[width=3.5in]{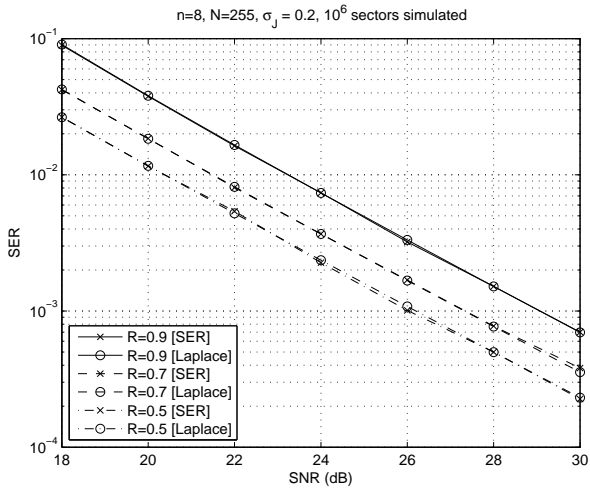}
\end{center}
\caption{The rate of exponential decay of Reed-Solomon SER with SNR does not depend on the code rate.}
\label{fig:gj_rs_rate}
\end{figure}

In Figure \ref{fig:gj_rs_rate} we compare numerical simulations of RS codes of various rates with expression (\ref{eq:erf}) which
is valid beyond the high rate approximation used to derive (\ref{eq:rs_ef_th}). We find a good agreement with our theoretical prediction:
the rate of the exponential decay does not depend on either the block size or the code rate, the position of the error floor changes slowly with
the code rate.

We conclude that non-interleaved Reed-Solomon codes applied to channel (\ref{chan:glob_jit})
exhibit an error floor: the probability of sector error decays exponentially as a function of $SNR$.
Moreover, the rate of the exponential decay does not depend on either the code rate
or the block size. The position of the error floor varies as a logarithm of the code rate.
Non-interleaved Reed-Solomon codes are thus not suitable for use in a probe storage system that suffers from
global positioning errors.

The above discussion suggests that the performance of Reed-Solomon codes can be improved with interleaving: by spreading
the codewords over multiple time samples, the effect of occasional strong jitter leading to $p<p_c(\tau)$ can be mitigated.
We will discuss the complexity of this solution
in Section \ref{sec:goodcodes}.

But first, we will address the following foundational question: do good codes for a global jitter channel exist in principle?

\section{Shannon Capacity of the Global Jitter Channel.}\label{sec:cap}

Shannon's capacity $C$ is one of the most important measures of quality for any communication channel. According to Claude Shannon's 1948 Channel Coding
Theorem \cite{Shannon2001}, there exists no error correction code of rate $R$ which would achieve an arbitrarily small probability of error for a channel with capacity smaller than code
rate, $C<R$.
The 'positive' part of Shannon's theorem states that for any $R < C$ and
for any $\epsilon > 0$ then there exists a block code $C$ with block size $M(\epsilon)$ and rate less than or equal to $R$
and a decoding algorithm such that the maximal probability of block error is less than $\epsilon$.
Thus by computing Shannon's capacity for the global jitter channel we will establish an upper bound on the rate of good error correction codes for this channel.

The definition of capacity rests on the notion of mutual information, see \cite{David2003} for a review of fundamental notions of information theory.
The mutual information between a discrete channel input $\mathbf{a}$ taking values in $\Omega_{a}$ and a continuous output ensemble $\mathbf{r}$ taking
values in $\Omega_r$ is defined as follows:
\begin{eqnarray}
I(R;A) = \sum_{\av\in \Omega_a}\Pr_A(\av)\int\limits_{\Omega_r} d\Pr_{R |A} (\rv|\av)\log_2\frac{d\Pr_{R|A}(\rv|\av)}{d\Pr_R(\rv)},\label{eq:mi}
\end{eqnarray}
where
$$
\frac{d\Pr_{R|A}(\rv|\av)}{d\Pr_R(\rv)}
$$
is Radon-Nykodim derivative of the conditional probability measure $\Pr_{R|A}$ with respect to the marginal probability measure $\Pr_{R}$.
If the probability densities $\rho_{R\mid A} (\rv \mid \av)$ and $\rho_R(\rv)$ of the probability measures $\Pr_{R|A}$ and $\Pr_R$ exist, the Radon-Nykodim
derivative is simply the ratio of densities:
$$
\frac{d\Pr_{R|A}(\rv|\av)}{d\Pr_R(\rv)}=\frac{\rho_{R|A}(\rv|\av)}{\rho_R(\rv)}.
$$

Recall that for the global jitter channel, $\av$ is an $N$-bit data string and $\rv$ is a string of $N$ real valued signals generated by the probe array at a given
sample time.

Shannon capacity is defined as the maximal mutual information over all input probability distributions per bit of input:
\begin{eqnarray}
C = \max_{P_A} \frac{1}{N} I(R;A)\label{eq:sh_cap}
\end{eqnarray}

As it is easy to see from the definition, $0\leq C\leq 1$ (bit). In particular, $C=1$ for noiseless channels. No information can be communicated
over the channel with $C=0$ in finite time.

For a complicated channel where the maximisation over $\Pr_A$ is difficult to perform it is common to study a weaker form of capacity:
\begin{eqnarray}
C_{i.u.d.} = \frac{1}{N}I(A;R)\bigg|_{\Pr_A=\Pr_{i.u.d.}}
\end{eqnarray}
where $\Pr_{i.u.d.}$ is the probability distribution for which channel inputs are chosen independently and with equal
probability. It is clear that
$$C_{i.u.d.}\leq C.$$
Therefore if we find that $C_{i.u.d.}$ is close to $1$, then $C$ is also close to one and we can be certain that there is a high rate ECC scheme which will achieve
low probability of error for the global jitter channel. From this point onwards we will be only concerned with $C_{i.u.d.}$ and
assume that $\Pr_A(\vect{a})=\frac{1}{2^N}$, i. e. that all $2^N$ inputs sequences
$\vect{a}=(a_1,\ldots,a_N)$ are sampled independently according to a uniform distribution.

The calculation of channel capacity is a notoriously difficult problem.
Analytically it can be evaluated for the very simplest channels only, such as binary symmetric, binary erasure or AWGN channels \cite{David2003}, Chapter $II$.
There exist efficient numerical algorithms for calculating capacity for channels with rapidly decaying correlations such as ISI channels, \cite{Arnold2001}.
Unfortunately, the numerical evaluation of capacity is not an option for channels with long range correlations such as the global jitter channel:
due to the strong correlations between $all$ signals received at the same time, the calculation of the $N$-dimensional integral in the right hand side of (\ref{eq:mi}) cannot
be reduced to a set of low dimensional problems which makes the
numerical evaluation of capacity extremely inefficient.

Fortunately, it turns out that capacity of channel (\ref{chan:glob_jit}) can be calculated asymptotically in the limit of large array size $N>>1$.
The simplification which allows capacity calculation in the large-$N$ limit is easy to understand: on the one hand, the received signals $\rv$ are independent
conditionally on the value of the amplitude $p=p(J)$. Conditionally on $p$, channel capacity is given by the well known expression for binary AWGN channels.
On  the other hand, the value of $p(J)$ can be extracted from the string of $N>>1$ received signals with relative accuracy of the order of $1/\sqrt{N}$ due to the
law of large numbers.

This simple argument suggests the following answer for capacity of the global jitter channel:
\begin{eqnarray}
\lim_{N\rightarrow\infty}\frac{1}{N}C_{i.u.d.} = \mathbb{E}_P\left[C_{AWGN}(p)\right]\label{eq:cap}
\end{eqnarray}
where $C_{AWGN}(p)$ is the capacity of the AWGN channel with fixed signal amplitude $p$ given by:
\begin{eqnarray}
C_{AWGN}(p) = 1 - \frac{1}{\sqrt{2\pi}}\int_{-\infty}^{+\infty}dx\left[ \exp\left(-\frac{x^2}{2}\right)\right.\nonumber\\\left.\times\log_2
\left(1 + \exp\left(fx-\frac{f^2}{2}\right)\right) \right]\label{eq:cap_awgn}
\end{eqnarray}
where $f=p/\sigma$,  and $\sigma$ is the standard deviation of additive white noise, see \cite{David2003}, Chapter II, for more details.
In what follows we will also use the asymptotic expansion of $C_{AWGN}$ valid in the limit of weak noise, $f>>1$:
\begin{eqnarray}
1-C_{AWGN}(p) = \frac{\sqrt{2\pi}}{f\ln(2)}\exp\left(-\frac{f^2}{8}\right)\left(1 + O\left(f^{-2}\right)\right).\label{eq:cap_asym}
\end{eqnarray}

As it turns out, the derivation of (\ref{eq:cap}) is very simple and relies on some basic properties of mutual
information which are both fundamental and intuitively obvious. Firstly, we notice that the mutual information for AWGN channel is maximized
for the uniform distribution of inputs. Then a simple rearrangement of terms in (\ref{eq:mi}) gives the following:
\bea
I(R;A)-N\mathbb{E}_P(C_{AWGN}(p))=I(P;R)-I(P;(A,R)),\label{eq:temp1}
\eea
where $I(P;R)$ is the mutual information between signal amplitude degraded by jitter and the received signal, $I(P;(A,R))$ is the mutual information
between signal amplitude and the joint ensemble of channel input and output. Clearly,
\bea
I(P;(A,R))\geq I(P;R).\label{eq:ineq1}
\eea
(The information we learn about $P$ from observing $A$ {\it and} $R$ must be greater or equal to the information about $P$ contained in $R$.)
In case the above argument fails to convince a rigorous-minded reader, here is the proof based on Jensen inequality \cite{Durret2004}:
$$
I(P;R)-I(P;(A,R))=\mathbf{E}_{(P,R,A)}\log \frac{d\Pr_{(P,R)}}{d\Pr_{(P,R)\mid A}} \frac{d\Pr_{R\mid A}}{d\Pr_R}
$$
$$
\stackrel{Jensen}{\leq} \log\mathbf{E}_{(P,R,A)} \frac{d\Pr_{(P,R)}}{d\Pr_{(P,R)\mid A}} \frac{d\Pr_{R\mid A}}{d\Pr_R}
$$
$$
=\log \int_{\Omega_p\times \Omega_r}d\Pr(P,R)=\log 1=0.
$$
Using (\ref{eq:ineq1}), relation (\ref{eq:temp1}) leads to the following inequality:
$$
0\leq \mathbb{E}_P(C_{AWGN}(p))-\lim_{N\rightarrow\infty}C_{i.u.d} \leq \lim_{N\rightarrow \infty} \frac{1}{N}I(P;(A,R)).
$$
Therefore, to verify (\ref{eq:cap}) it remains to show that
\bea
\lim_{N\rightarrow \infty} \frac{1}{N}I(P;(A,R))=0.\label{eq:temp22}
\eea
Intuitively, the validity of the above claim is fairly obvious: imagine for example that random variable $P$ is represented by $m$-bit numbers.
Then the mutual information $I(P;(A,R))$ cannot exceed $m$ bits and the limit in the right hand side (\ref{eq:temp22}) is trivially zero.
For the proof of (\ref{eq:temp22}) in full generality, the reader is referred to Appendix \ref{ap:limit}.

\begin{figure}
\begin{center}
\includegraphics[width=3.5in]{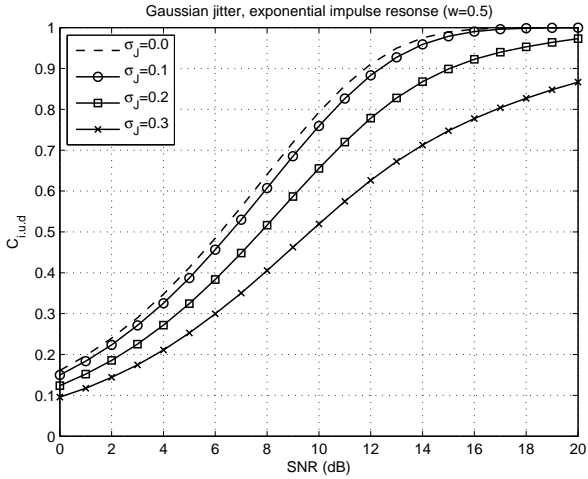}
\caption{Capacity (i.u.d) for the global jitter channel}
\label{fig:capacity}
\end{center}
\end{figure}

The problem of computing capacity of the global jitter channel for large values of $N$ is solved in principle: the
right hand side of (\ref{eq:cap}) is a finite-dimensional integral which depends on the probability distribution of $p(J)$. In particular,
it is well suited for numerical study. In Figure \ref{fig:capacity} the capacity of the probe storage
channel suffering a global Gaussian jitter is shown for various values of jitter strength $\sigma_J$. For $\sigma_J=0$ the channel is 
equivalent to the binary AWGN channel with signal amplitude equal to one. As seen from the plot, this channel has the highest capacity. For a fixed SNR,
the capacity of global jitter decreases as $\sigma_J$ increases. For example at the SNR point corresponding to $C_{AWGN}=0.9$ bits, the capacity
of global jitter channel with $\sigma_J=0.3$ is $C_{i.u.d}=0.62$ bits, which rules out the use of high rate linear error correction codes for this channel.
The good news is that capacity seems to approach $1$ bit per channel symbol in the limit of high SNR even in the presence of global jitter. According to Shannon's theorem this
means that by reducing channel noise one can read and write information reliably using large parallel probe arrays at a small redundancy cost.

\begin{figure}
\begin{center}
\includegraphics[width=3.5in]{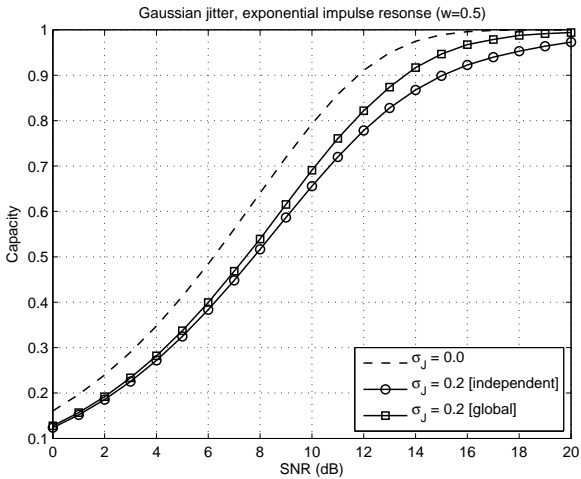}
\caption{Comparison of capacity (i.u.d) for the global jitter channel against the reference channel}
\label{fig:capacity_compare}
\end{center}
\end{figure}

Moreover, the global jitter channel can actually have a larger capacity than a collection of independent probes subject to individual jitter
distortions of the same strength: In Figure \ref{fig:capacity_compare} the
capacity of the global jitter channel is compared against the reference channel where each probes
suffers an independent Gaussian jitter distortion of the same strength  for which the i.u.d capacity can be easily computed numerically.
We observe that at low SNR the reference channel has the same capacity than the global jitter channel but at high SNR the global jitter channel actually has a larger capacity!
We can explain this phenomenon by the fact that strong correlations between position jitters across the whole array can be used to extract extra information
about the hidden parameters of the channel. This extra information can be used for example to build better signal detection algorithms, as discussed in Section \ref{sec:opdet}.

To complete our investigation of $C_{i.u.d.}$ we still need to: (i) confirm the numerical observation that capacity approaches $1$ bit in the limit of low channel noise; (ii) estimate
the corresponding rate of convergence.
We will solve these problems for the exponential impulse response assuming weak additive noise and weak jitter, i. e.
\bea
\sigma<<1,~\sigma_J<<W.
\eea
While both of these conditions are reasonable in the context of applications, the weakness of jitter leads to a significant technical simplification of the argument
given below.
Our starting point is the following bound on the capacity of the AWGN channel:
\bea
1-U\frac{\sigma}{p}e^{-\frac{p^2}{8\sigma^2}}\geq C_{AWGN}(p)\geq 1-L e^{-\frac{p^2}{8\sigma^2}},\label{eq:bound2}
\eea
for some positive constants $U$ and $L$. Eq. (\ref{eq:bound2}) results from the straightforward yet tedious analysis of (\ref{eq:cap_awgn}).
Averaging the above inequality over $p$ we obtain the following bound for the capacity of the global jitter channel:
\bea
\lim_{N\rightarrow \infty} C_{i.u.d.}\geq 1-L'\nonumber\\ \times \frac{W}{\sigma_J}\left(2^{\frac{W^2}{4\sigma_J^2}-1} 
\Gamma\left(\frac{W^2}{2\sigma_J^2}\right) \sigma^{\frac{W^2}{2\sigma^2_J}}+\sqrt{\pi} e^{-\frac{1}{2e^2\sigma^2}}  \right)
\label{eq:caplb}\\
\lim_{N\rightarrow \infty} C_{i.u.d.}\leq 1-U'\frac{W}{\sigma_J}\frac{\sigma^{\frac{W^2}{2\sigma_J^2}}}{\sqrt{\log(1/8\sigma^2)}}I\left(\frac{W}{2\sigma_J}\right),
\label{eq:capub}
\eea
 where $L'$ and $U'$ are positive constants, $\Gamma(x)$ is the $\Gamma$-function,
$$
I(z)=8^{\frac{z-1}{2}}\int_{0}^{1/e}dx x^{z-2} e^{-x^2}.
$$
As the right hand side of the lower bound (\ref{eq:caplb}) approaches $1$ bit in the low noise limit $\sigma=0$, we conclude that
$$
\lim_{\sigma \rightarrow 0}\lim_{N\rightarrow \infty} C_{i.u.d.}=1,
$$
which confirms the results of numerical simulations.
The upper bound (\ref{eq:capub}) shows that the convergence to one is not faster than the power law $\sigma^{\frac{W^2}{2\sigma_J^2}}$, which
is much slower than the exponential convergence of $C_{AWGN}(p) \sim 1-Ae^{-p^2/8\sigma^2}$ to one - only a very 'clean'  global jitter
channel will have capacity close to the capacity of AWGN channel.

Finally, comparing the lower capacity bound with the upper
bound we see that with logarithmic precision the convergence of $C_{i.u.d.}$ for large $N$ is indeed given by the power law:
\bea
1-C_{i.u.d.}\sim Const \cdot \sigma^{\frac{W^2}{2\sigma_J^2}}.\label{eq:pl}
\eea
Therefore, the capacity approaches $1$ bit in exactly the same way as the Reed-Solomon sector error rate vanishes in the limit of zero additive noise, see (\ref{eq:rs_ef_th}).
The power law approach of the channel characteristics to their noiseless limiting values seems to be a feature of the global jitter channel. We will encounter
the law (\ref{eq:pl}) again in the next Section, when we compute the Gallager's coding bound for the global jitter channel.

\section{The random coding bound for the global jitter channel.}\label{sec:rcb}
According to the results of the previous Section, the capacity of the global jitter is positive and even approaches $1$ bit in the limit of low noise. Therefore, it follows from
Shannon's theorem that there are error correction codes with rate up to capacity $C$ which can be used to transmit information over the channel with vanishingly small probability of error.
Yet as we have seen in Section \ref{sec:rs}, non-interleaved Reed-Solomon codes
exhibit an error floor for any block size and any rate.

Therefore, we are led to the following question: are there $any$ error correction block codes (linear or non-linear) which yield a small probability of block error
if the code block coincides with the instantaneous output of the probe array?

Let $\mathfrak{C}$ be a code with rate $R$ and block size $N$ (bits) which is also the number
of probes in the array. Let $\Pr(SE\mid \mathfrak{C})$ be the probability
of block decoding error for this code.

This probability is difficult to compute for any non-trivial code.
However, we can use the idea of  Gallager \cite{Gallager1965} and calculate the $average$
probability $\overline{\Pr}(SE\mid R,N)$ of block error under the maximum likelihood decoding,
where the average is taken over all random codes with a given rate $R$ and block size $N$.
Given $\overline{\Pr} (SE\mid R,N)$ there must exist a code $\mathfrak{C}_0$ with rate $R$ and block size $N$ such that
$$
\Pr(SE\mid \mathfrak{C}_0)\leq \overline{\Pr} (SE\mid R,N).
$$
Therefore, if we find that $\overline{\Pr}(SE\mid R,N)$ is sufficiently low (e. g. $10^{-12}$)
for a desired  code rate $R$, we will know that there are codes which can be used to
correct errors for information communicated over the global jitter channel.

To perform the calculation of the random coding bound we need to define the space of random codes and a probability
measure on this space. Following Gallager, we will construct a
random binary code by picking $K=NR$ binary codewords from the $N$-dimensional binary space $\{0,1\}^N$
independently and uniformly (Gallager ensemble).

Using the sum rule, $\Pr(SE\mid \mathfrak{C})$ can be re-written as follows:
\begin{eqnarray*}
\Pr (SE\mid \mathfrak{C}) = \int_0^1d\mu_P(p) \Pr(SE \mid \mathfrak{C},p )
\end{eqnarray*}
where $\Pr(SE \mid \mathfrak{C},p )$ is the probability of block
decoding error for the binary AWGN channel with a fixed signal amplitude $p$.
Therefore,
\bea
\overline{\Pr}(SE\mid R,N)=\int_0^1d\mu_P(p) \overline{\Pr} (SE \mid p, R, N ),\label{eq:pe_gj}
\eea
where $\overline{\Pr}(Se \mid p,R,N)$ is the random coding bound for the binary AWGN channel
with a fixed signal amplitude $p$. The following set of results can be easily extracted from the original
Gallager's paper \cite{Gallager1965}:
\bea
 \overline{\Pr} (SE \mid p, R, N )\leq e^{-NE(R,p)},
\eea
where $E(R,p)$ is the error exponent for the binary AWGN channel.
In what follows we do not need an explicit expression for the error exponent,
but its basic properties listed below will be important to us:
\bea
E(R,p)\equiv 0 \mbox{ for } R\geq C_{AWGN}(p), \label{eq:null}\\
E(R,p)>0 \mbox{ for }  R< C_{AWGN}(p),\label{eq:pos}\\
\frac{\partial E}{\partial R}(R,p)\mid_{R=C_{AWGN}(p)}=0,\label{eq:min}\\
\frac{\partial^2 E}{\partial R^2}(R,p)\mid_{R=C_{AWGN}(p)}>0.\label{eq:conv}
\eea

Using the notion of the error exponent for AWGN channel, we can re-write
the random coding bound for the global jitter channel as follows:
\begin{eqnarray}
\overline{\Pr}(SE\mid R,N) \leq \int_0^1d\mu_P(p)\exp\left[-NE(R,p)\right].\label{eq:pe_gj_ee}
\end{eqnarray}
Let $p_c$ be the unique solution to
\begin{eqnarray}
C_{AWGN}(p) = R
\end{eqnarray}
Then (\ref{eq:pe_gj_ee}) can be rewritten as:
\begin{eqnarray}
\overline{\Pr}(SE\mid R,N) \leq \int_0^{p_c}d\mu_P(p)\exp\left[-NE(R,p)\right]\nonumber\\
 + \int_{p_c}^{1}d\mu_P(p)\exp\left[-NE(R,p)\right]
\end{eqnarray}
The crucial observation is that for $p\leq p_c$, $C_{AWGN}(p) \leq R$. As a result, for $p\leq p_c$, $E(R,p)\equiv 0$ due
to (\ref{eq:null}). Therefore the average probability of decoding error for the global jitter channel is bounded:
\begin{eqnarray}
\overline{\Pr}(SE\mid R,N) \leq \int_0^{p_c}d\mu_P(p)\nonumber\\ + \int_{p_c}^{1}d\mu_P(p)\exp\left[-NE(R,p)\right]\label{eq:pe_gj_2int}
\end{eqnarray}

The second integral in the right hand side of (\ref{eq:pe_gj_2int}) can be evaluated for $N>>1$ using Laplace method:
due to (\ref{eq:pos}, \ref{eq:min}) the main contribution to the integral comes from a small neighbourhood of $p_c$.
It follows from (\ref{eq:min},\ref{eq:conv}) that as a function
of $p$, $E(R,p)$ has a non-degenerate critical point at  $p=p_c$.
A calculation based on the above points yields:
\begin{eqnarray*}
\overline{\Pr}(SE\mid R,N) \leq \Pr(p<p_c) \\+ \rho_P(p_c)\sqrt{\frac{\pi}{N\partial_R^2E(R,p_c)\partial_pC(p_c)^2}}+O(N^{-1}).
\end{eqnarray*}

It is remarkable that in the limit $N\rightarrow\infty$ the probability of block decoding error is bounded by a function independent of block size:
\begin{eqnarray}
\lim_{N\rightarrow \infty} \overline{\Pr} (SE\mid N,R) \leq \Pr(p<p_c)\label{eq:ps_rcb}
\end{eqnarray}
Therefore, the random coding bound for the global jitter channel does not vanish in the limit
of the large block size!

\begin{figure}
\begin{center}
\includegraphics[width=3.5in]{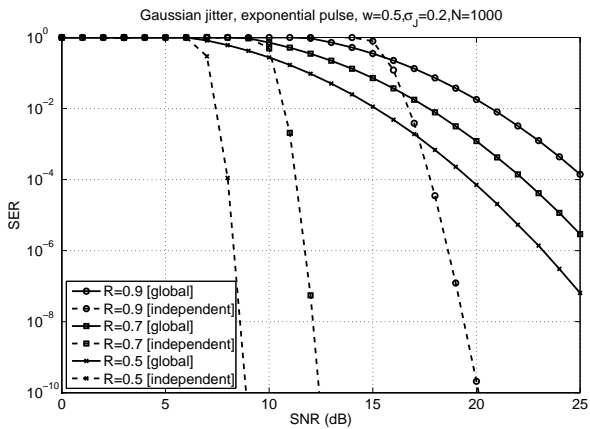}
\caption{Random coding bound for the probe storage channel with independent and global jitter for various different rates}
\label{fig:rcb_ps}
\end{center}
\end{figure}

In Figure \ref{fig:rcb_ps} the random coding bound  (\ref{eq:ps_rcb}) is shown for various code rates for a global
Gaussian positioning error with strength $\sigma_J=0.2$. The random coding bound for a channel where each probe in an array of
$N=1000$ suffers an independent Gaussian positioning error of the same strength is also given for the same code rates. It is observed that the
probability of decoding error (SER) is vastly worse for the case of global positioning errors. For the independent
channel the SER decay with a \textit{waterfall} (super-exponential) shape whereas for the global jitter channel the SER decays exponentially with SNR, i. e. exhibits an error floor.

\begin{figure}
\begin{center}
\includegraphics[width=3.5in]{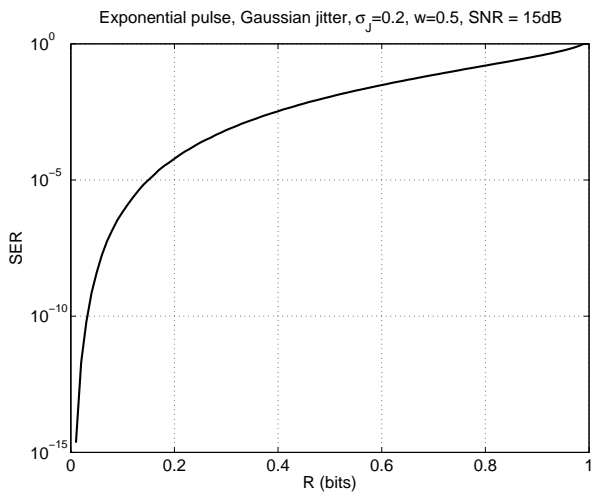}
\caption{Behaviour of the random coding bound as a function of rate for the probe storage channel suffering a global Gaussian positioning error}
\label{fig:rcb_rate}
\end{center}
\end{figure}

The presence of the error floor in the random coding bound can be demonstrated
analytically for Gaussian impulse response and Gaussian jitter in the limit of high SNR and high rate codes.
The cumulative distribution function $\Pr(p\leq p_c)$ is given by:
\begin{eqnarray}
\Pr(p\leq p_c) &=& \frac{2}{\sqrt{2\pi\sigma_J^2}}\int_{J_c}^{\infty}dJ\exp\left[-\frac{J^2}{2\sigma_J^2}\right]\nonumber\\&\leq& \exp\left[-\frac{J_c^2}{2\sigma_J^2}\right]
\end{eqnarray}
Where $J_c=W\sqrt{\ln(1/p_c)}$ is the positive value of jitter that results in signal loss $p$.
Recall from (\ref{eq:ex_imp}) that $W$ is parameter related to pulse width.  Thus the average probability of decoding error is bounded by:
\begin{eqnarray}
\lim_{N\rightarrow \infty} \overline{\Pr}(SE\mid R,N) \leq p_c^{\left(\frac{W^2}{2\sigma_J^2}\right)}\label{eq:rcb_ef1}
\end{eqnarray}
Furthermore in the limit of weak noise $\sigma\rightarrow 0$ and high rate $R\rightarrow 1$, it is possible to use the
asymptotic expansion for AWGN capacity $C_{AWGN}(p)$ given by (\ref{eq:cap_asym}) to derive an expression for $p_c$ to the logarithmic precision:
\begin{eqnarray}
p_c \approx \sigma\sqrt{8\ln\left(1/(1-R)\right)}
\end{eqnarray}
Thus for high-rate codes the probability of decoding error can be approximately bounded:
\begin{eqnarray}
\lim_{N\rightarrow \infty} \overline{\Pr}(SE\mid R,N) \lesssim C(\sigma_J,w,R)\; \sigma^{\left(\frac{W^2}{2\sigma_J^2}\right)}
\label{eq:rcb_ef2}
\end{eqnarray}
where $C(\sigma_J,w,R)$ is the $\sigma$-independent constant:
\begin{eqnarray}
C(\sigma_J,w,R) = \left(-8\ln\left(1-R\right)\right)^{\left(\frac{W^2}{4\sigma_J^2}\right)}
\end{eqnarray}
We conclude the average probability of decoding error in the limit of weak noise and high code rate is asymptotically independent of the code block size, exhibits
an error floor with an exponent independent of the code rate $R$ and an amplitude with depends on the rate via $log(1-R)$. An identical behaviour has been
observed for Reed-Solomon codes in Section \ref{sec:rs}.
Similar conclusions concerning the random coding bound can be reached for low code rates as well, see
Fig. \ref{fig:rcb_rate} where the bound (\ref{eq:ps_rcb}) is shown as a function of rate $R$ for a fixed SNR.
Note that the average probability of block error approaches zero only in the limit of zero code rate.

The fact that the random coding bound exhibits an error floor behavior identical to that of RS code suggests to us that $all$
non-interleaved error correction codes suffer identical performance degradation due to global jitter. This suggestion
is confirmed in the following Section.

\section{The non-existence of non-interleaved error correction codes for the global jitter channel.}\label{sec:fano}
The existence of an $N$-independent error floor in the average probability of block error has the following simple
explanation: provided the strength of additive noise is positive and $R>0$, there is an $N$-independent
critical value of jitter $p_c(R)$
beyond which the conditional channel capacity $C_{AWGN}(p)$ is smaller than the code rate $R$. Therefore,
according to the negative part of Shannon's theorem, information transmission without errors
is impossible with $N$-independent probability $\Pr(p<p_c(R))$.

The above consideration can be turned into a rigorous argument which shows that non-interleaved encoding
of the large array's outputs cannot ensure error free information retrieval {\it no matter what error correction
code with positive rate is used}:

Let $\mathfrak{C}(R,N)$ be a block code with rate $R$ and block size $N$. Clearly,

\bea
\Pr(SE\mid\mathfrak{C}(R,N))\geq \int_{0}^{p_c} d\mu(p) \Pr(SE\mid p,\mathfrak{C}(R,N)).
\eea
Recall that $p_c$ is the unique solution to $R=C_{AWGN}(p)$.
Notice that for every $p$ in the region of integration, $R>C_{AWGN}(p)$.
By Fano's inequality \cite{Fano1961},
\bea
\Pr(SE \mid p, \mathfrak{C}(R,N))\geq \frac{R-C_{AWGN}(p)}{R}-\frac{1}{RN}.
\eea
Therefore,
\bea
\Pr(SE\mid\mathfrak{C}(R,N))&\geq& \int_{0}^{p_c}d\mu(p)\left(1-\frac{C_{AWGN}(p)}{R}\right)\nonumber\\&-&\frac{1}{RN}\Pr(p\leq p_c).\label{eq:pre1}
\eea
Therefore, the probability of block error is bounded below by a constant asymptotically independent of the block size.

Integrating (\ref{eq:pre1}) by parts it is easy to show that
\bea
\lim_{N\rightarrow \infty}\Pr(SE\mid \mathfrak{C}(R,N))\geq \int_{0}^R \frac{dc}{R} \Pr(C_{AWGN}(p)\leq c),\label{eq:pre2}
\eea
where $$\Pr(C_{AWGN}(p)<c)=\int_0^1 d\mu(p)\chi(C_{AWGN}(p)<c).$$
Expression (\ref{eq:pre2}) proves that the probability of block error is bounded away from zero by an $N$-independent constant for
$any$ non-interleaved error correction code with a positive rate $R$.

Therefore, error-free information storage using large arrays of probes is impossible without interleaving error correction codes between multiple
array outputs.

It is worth noting that there is a counterpart of Fano's inequality for the global jitter channel - a non-trivial fact given strong correlations between all probe channels within
the array. Using the fact that $\lim_{N\rightarrow \infty}C_{i.u.d.}=\mathbb{E}_p(C_{AWGN}(p))$ which we established in Section \ref{sec:cap} we can derive the following weaker bound from the bound (\ref{eq:pre2}):
\bea
\lim_{N\rightarrow \infty}\Pr(SE\mid \mathfrak{C}(R,N))\geq 1-\frac{\lim_{N\rightarrow \infty}C_{i.u.d.}}{R},\label{eq:fano}
\eea
which shows the impossibility of error-free information transmission over the global jitter channel using non-interleaved codes with $R>\lim_{N\rightarrow \infty}C_{i.u.d.}$.

Finally, let us show that any non-interleaved code used to encode information transmitted over Gaussian global jitter exhibits an error floor identical to Reed-Solomon
error floor (\ref{eq:rsup2}, \ref{eq:rs_ef_th}) or the error floor in the random coding bound (\ref{eq:rcb_ef1}, \ref{eq:rcb_ef2}).
Note that the capacity of AWGN channel $C_{AWGN}(p)$ is a function of $p/\sigma$ only,
$$
C(p)=F\left(\frac{p}{\sigma}\right),
$$
see (\ref{eq:cap_awgn}). Substituting expression (\ref{eq:pdf}) for the probability measure $d\mu_P(p)$ for Gaussian jitter into the integral in the right hand side of (\ref{eq:pre1})
and changing the integration variable $p=\sigma x$ we arrive at the following result:
\bea
lim_{N\rightarrow \infty} \Pr(SE\mid\mathfrak{C}(R,N))\nonumber\\\geq\frac{\sigma^\gamma}{\sqrt{2\pi \sigma_J^2}}
\int_0^{x_c}dx \frac{x^{\gamma-1}}{\sqrt{log\frac{x_c}{x}+log\frac{1}{\sigma x_c}}}\left( 1-\frac{F(x)}{R} \right)\label{eq:temp3}
\eea
where $\gamma=\frac{W^2}{2\sigma_J^2}$ and $x_c$ is the unique positive solution to the equation
\bea
F(x_c)=R.
\eea
Notice that $x_c\sigma=p_c<1$.
The following inequality is valid provided the additive noise is weak: If
$$
log\frac{1}{x_c\sigma}>1/2:
$$
and for any $x:~0<x\leq x_c $,
\bea
\frac{1}{\sqrt{log\frac{x_c}{x}+log\frac{1}{\sigma x_c}}}\geq \frac{1}{\sqrt{log\frac{1}{x_c\sigma}}}\frac{x}{x_c}
\eea
Using this inequality in (\ref{eq:temp3}) we conclude the following:
\bea
lim_{N\rightarrow \infty} \Pr(SE\mid\mathfrak{C}(R,N))\geq I(R)\frac{\sigma^{\frac{W^2}{2\sigma_J^2}}}{\sqrt{2\pi \sigma_J^2\log\frac{1}{x_c\sigma}}},\label{eq:errf}
\eea
where
$$
I(R)=\int_0^{x_c}\frac{dx}{x_c} x^{\gamma}\left( 1-\frac{F(x)}{R} \right)
$$
is $\sigma$-independent function of $R$ and $\gamma=\frac{W^2}{2\sigma_J^2}$.

We conclude that $any$ non-interleaved error correction code with rate $R$ and a large block size used to transmit information over the global
jitter channel exhibits an exponential error floor: $log \Pr(SE)$ is a linear function of signal-to-noise ratio with exponent
\bea
\gamma=\frac{W^2}{2\sigma_J^2}.
\eea
The position of the error floor is determined by the function $I(R)$, which is a slow function of code rate, but we do not study it here.
Note that the upper bound on the error floor we derived using elementary tools only (Fano's inequality) coincides with the error floor observed
directly for Reed-Solomon codes and the random coding bound for the global jitter channel.

\section{On good codes for the global jitter channel.}\label{sec:goodcodes}
In the two previous sections we established the impossibility of error-free information retrieval for large parallel arrays of probes subject to global jitter
using non-interleaved error correction codes of $any$ positive code rate.

We also established in Section \ref{sec:cap} that the capacity $C_{i.u.d.}$ of global jitter channel is positive and approaches $1$ bit as the additive noise strength $\sigma$
goes to $0$.
Therefore, by Shannon's theorem there must exist families of error correction codes with rates $0<R<C_{i.u.d.}$ which ensure vanishingly small probability of block error in the limit
of large block sizes.

Can we say anything about the structure of these  codes? We understand that non-interleaved codes cannot perform well on the global jitter channel due to rare
strong jitter fluctuations leading to small effective AWGN amplitude $p \lesssim \sigma$. Due to this fluctuations $N$ bits of information get lost regardless
of error correction code used.

To avoid this we must spread information over many time slices by encoding $blocks$ of array outputs. Therefore good error correction codes for the global jitter
channel must be interleaved.

Let us estimate the depth of interleaving and the block size of the corresponding codes.
Let ${\rv}_{t\in \mathbb{Z}}$ be the time series of array outputs. Recall that each output is an $N$-dimensional vector, where $N$ is the number of probes in the
array. For a feedback-loop based positioning system used in Millipede, jitter random variables $J_t$ are correlated in time. Let $L$ be the correlation
length measured in the number of sampling periods.
Our previous discussion on the influence of correlations on the performance of error correction codes can be summed up as follows: the probability of decoding
error of the block code of rate $R<C_{i.u.d.}$ and block size $B$ approaches zero as the number of independent groups of samples $\frac{B}{NL}$ goes to infinity.
In other words, if $NL$ strongly correlated samples are treated as a single symbol, we get the usual statement of Shannon's theorem for memoryless channels: the probability of
block error goes to zero as the number of symbols per block goes to infinity.

The minimal block size of high rate error correction codes used in modern storage devices which achieve the probability of error of the order
of $10^{-10}$ is of the order of $10^3$ bits. Clearly, this is the lower bound on the interleaving depth.
Therefore we can estimate the block size of the interleaved error correction code for probe storage as
\bea
B \sim 10^3 NL
\eea

Thus for a probe storage device with $N \sim 10^3$ probes in the array the error correction block size should be of the order
$$
B \gtrsim 10^6 \mbox{ bits }.
$$

The codes of similar sizes ($5\cdot 10^5$ bits) are already used in optical storage (Blu-Ray disks), however we thought it unusual to have discovered the need
for large block sizes for a non-removable media storage device.

\section{Conclusion}\label{sec:concl}
In this paper we introduced and analyzed a simple model of the massively parallel
probe storage channel suffering from global positioning errors. This model can be viewed
as an example of communication channel consisting of $N>>1$ parallel sub-channels all of which
are strongly correlated via a common distortion event.

We solved the problem of optimal signal detection for the global jitter channel.
Namely we found a detection algorithm of $O(NlogN)$ complexity such that the estimated sequence
converges bitwise to the maximum a posteriori estimate in the limit $N\rightarrow \infty$.
This is not an entirely trivial result, as in general, the complexity of optimal detection grows exponentially
with memory length (for example, the optimal MAP detector for the channel with inter-symbol interference of length $I$ and
AWG noise has $2^I$ states). For the channel at hand the detection algorithm simplifies
due to statistical independence of the channel outputs conditional on the value of jitter. This value can be found
by analyzing all channel outputs using the law of large numbers.

We analyzed the performance of Reed-Solomon codes applied to the global jitter channel without interleaving
channel inputs corresponding to different moments of time. We discovered both numerically and theoretically
that in the limit of large probe array, any non-interleaved Reed-Solomon code will exhibit an error floor
the position of which is independent of the code's block size $N$ and only weakly dependent on the code rate.
This is a surprising result, as the phenomenon of the error floor is usually associated with sub-optimality of the 
decoding algorithm (e. g. belief propagation) rather than specifics of the channel. For the case of global jitter the origin
of the error floor can be traced back to rare instances of strong jitter $J$ such that the signal amplitude $p(J)$
becomes of the same order as channel noise $\sigma$.

Motivated by these findings we addressed the following question: are there any good error correction codes for probe storage channels?
The answer turned out to be two-fold: firstly, we calculated the capacity $C_{i.u.d.}$ of the probe storage channel and discovered
that it does approach one bit per channel symbol in the limit of weak channel noise $\sigma$, albeit very slowly: $1-C_{i.u.d.}$ approaches zero as $\sigma^\gamma$,
where the exponent $\gamma$ is equal to the error floor exponent for Reed-Solomon codes.
However, we also found that Gallager's random coding bound for channel encoding which does not interleave between different time slices exhibits exactly the same error
floor as Reed-Solomon codes! Moreover, an application of Fano inequality allowed us to prove that any non-interleaved error correction code
applied to a massively parallel probe storage channel
exhibits an error floor behaviour which is at least as bad as the Reed-Solomon error floor.

With the benefit of hindsight, the appearance of the universal error floor in Reed-Solomon block error rate, Shannon
capacity, Gallager's random coding bound and Fano's low bound on sector error rate for an arbitrary fixed code can be
related to the statistics of large jitter fluctuations: if $p(J)<< \sigma$
then regardless of the code used
regions of confuse-ability (a hypersphere of radius $\sigma$) for any two codewords intersect and all information and parity is lost!

In this sense the
global jitter channel for weak channel noise can be viewed as an effective block-wise erasure channel:
either $N$ bits are detected with no errors, or all $N$ bits are lost. For such a channel, it is clearly
impossible to improve performance by increasing the number of probes in the array if the encoding block size
is kept equal to $N$. In order to reach Shannon's limit for the global jitter channel it is
necessary to spread information between many outputs of the probe array thus mitigating the effects
of occasional strong jitter. We estimate the necessary block size of good error correction codes
for probe storage channel to be of the order of $10^6$ - an unusually large number for a non-removable storage device.

The mathematical results reported in the paper have severe practical implication for probe storage. Either the positioning system must be
accurate enough so that global jitter is made very weak or we must interleave the error correction code
in the time direction. The former option poses a huge engineering challenge: for indentation sizes of the order of $10$ nm, our
results suggest that the precision of the positioning system must be significantly better than $1$ nm for the effects of global
jitter to become insignificant. The latter option also presents significant implementation difficulties especially since
jitter is known to be strongly correlated in the time direction and as a result we must
interleave very deeply to overcome its effects.

Due to the universal nature of information theoretic performance
limits we discovered, we see no easy 'engineering' way around the problem of global jitter. For example one can
attempt to re-read the information recovered by the array of probes following a strong jitter event. This is similar
to the way off-track errors are dealt with in magnetic hard drives.  Our results
mean however that the throughput of such a system will be severely degraded due to a large number of necessary re-reads.

From a more theoretical point of view, we developed a general approach for solving the problem of performance evaluation of
channels with long correlations: If the correlations are
due to a small number of hidden 'correlation' parameters (e. g. global jitter), we can calculate the conditional capacity, the random coding bound, etc.
using well known expressions for memoryless channels and then average over the hidden parameters to obtain the unconditional quantities.
The universality of our answers can be explained by the fact that for a large number of sub-channels $N>>1$ only the values
of hidden parameters dictated by the law of large numbers are needed to remove the conditioning.

Given the general nature of our analysis, we hope that results reported in the present paper can be applied to other communications channels
with long range correlations between channel outputs. Relevant examples include high density magnetic storage
for which long (of the order of several sectors) correlations are present due to off-track errors; optical storage with
removable media for which correlations are generated by scratches on media surface; ultra
dense NAND flash memory (corresponding to $20$ nm and smaller transistor libraries)
where long correlations along both bit- and word-lines occur due to read and programme disturbs. An example coming from
digital communications is a multiple-input-multiple-output (MIMO) system of receivers and transmitters which is known to be
strongly affected by the spatial correlations in the noise vector, \cite{Salz1985}.

\appendices
\section{The derivation of bound (\ref{eq:bound})}\label{ap:bound}

Let $\hat{\hat{a}}_t^{(k)}$ and $\hat{a}_t^{(k)}$ be the $k$-th output of the genie detector (given by equation (\ref{eq:genie})) and the $k$-th output
of the LLN detector (given by equation (\ref{eq:det})) respectively. Then using the triangle inequality and the laws of
conditional probability it is possible to show:
\begin{eqnarray}
\left|\Pr\left(\hat{\hat{a}}_t^{(k)}\neq a_t^{(k)}\right)-\Pr\left(\hat{a}_t^{(k)}\neq a_t^{(k)}\right)\right| \nonumber \\ \leq
\int d\mu_p(\pi)\Pr\left( \hat{\hat{a}}_t^{(k)} \neq \hat{a}_t^{k} | p = \pi\right)\label{eq:tri}
\end{eqnarray}
The event corresponding to the output of the two detectors being different can be expressed as follows:
\begin{eqnarray}
\Pr\left(\hat{\hat{a}}_t^{(k)} \neq \hat{a}_t^{(k)} | p = \pi\right) = \nonumber\\ \Pr\left(r_t^{(k)} \in \left(\frac{p}{2},T_N\right)\bigg|\;
\left|\frac{p}{2}-T_N\right| > \epsilon, p = \pi \right)\nonumber \\\times \Pr\left( \left|\frac{p}{2}-T_N\right| > \epsilon
\bigg|\; p=\pi\right) \nonumber \\ +\Pr\left(r_t^{(k)} \in \left(\frac{p}{2},T_N\right)\bigg|\; \left|\frac{p}{2}-T_N\right| < \epsilon,
p = \pi \right)\nonumber \\\times \Pr\left( \left|\frac{p}{2}-T_N\right| < \epsilon \bigg|\; p=\pi\right)
\end{eqnarray}
We proceed to bound each term as follows upper-bound
\begin{eqnarray}
\Pr\left(r_t^{(k)} \in \left(\frac{p}{2},T_N\right)\bigg|\; \left|\frac{p}{2}-T_N\right| > \epsilon, p = \pi \right) \leq 1 \label{eq:up1}  \\
\Pr\left( \left|\frac{p}{2}-T_N\right| < \epsilon \bigg|\; p=\pi\right) \leq 1 \label{eq:up2}\\
 \Pr\left( \left|\frac{p}{2}-T_N\right| > \epsilon \bigg|\; p=\pi\right) \leq \frac{\operatorname{Var}
 \left(\frac{p}{2}-T_N|p=\pi\right)}{\epsilon^2} \label{eq:up3} \\
\Pr\left(r_t^{(k)} \in \left(\frac{p}{2},T_N\right)\bigg|\; \left|\frac{p}{2}-T_N\right| < \epsilon, p = \pi \right)\nonumber\\
\leq \frac{2\epsilon \; e^{-\frac{\left(\pi/2 - \epsilon\right)^2}{2\sigma^2}}}{\sqrt{2\pi\sigma^2}}\label{eq:up4}
\end{eqnarray}
Where (\ref{eq:up1}) and (\ref{eq:up2}) are trivial, (\ref{eq:up3}) is due to the central limit
theorem (and convergence of $T_N$ to $p/2$) and (\ref{eq:up4}) is a simple upper bound of the
corresponding Gaussian integral. Substituting the resulting upper bound into (\ref{eq:tri}) and integrating over $\pi$ we find:
\begin{eqnarray}
\left|\Pr\left(\hat{\hat{a}}_t^{(k)}\neq a_t^{(k)}\right)-\Pr\left(\hat{a}_t^{(k)}\neq a_t^{(k)}\right)\right| \nonumber\\ \leq
\frac{\mathbb{E}(p^2)/4 + \sigma^2}{N\epsilon^2} + \frac{2\epsilon}{\sqrt{2\pi\sigma^2}}\label{eq:final_bound}
\end{eqnarray}
We now minimise with respect to $\epsilon$ to obtain the tightest bound possible. The critical value of $\epsilon$ is given by:
\begin{eqnarray}
\epsilon_{min} = \left(\frac{(\mathbb{E}(p^2)/4 + \sigma^2)\sqrt{2\pi\sigma^2}}{N}\right)^{1/3}
\end{eqnarray}
 Substituting $\epsilon_{min}$ into (\ref{eq:final_bound}) we arrive at the bound (\ref{eq:bound}).

\section{The calculation of $\lim_{N\rightarrow \infty} \frac{1}{N}I(P;(A,R))$.}\label{ap:limit}
We will prove that $\lim_{N\rightarrow \infty} \frac{1}{N}I(P;(A,R))=0$ not just for Gaussian jitter, but
any distribution of $p\in (0,1)$ with finite differential entropy:
\bea
h=\mathbb{E}_P\left(log\frac{1}{p}\right)<\infty.
\eea

By definition, the mutual information between signal strength $P$ and the channel's input and output $(A,R)$ is:
\bea
I(P;\Rv)=\mathbb{E}_{P,A,R}\log\frac{d\Pr(A,R\mid P)}{d\Pr(A,R)},
\eea
Using the independence of signal strength $P$ and channel input $A$ the above expression can be re-written as follows:
\bea
I(P;\Rv)=\mathbb{E}_{P,A,R}\log\frac{d\Pr(R\mid A,P)}{d\Pr(R\mid A)}.
\eea
For Gaussian additive channel noise, the fraction under the sign of the logarithm takes the form
\bea
F(P,A,R)\equiv \frac{d\Pr(R\mid A,P)}{d\Pr(R\mid A)}\nonumber \\
=\int_0^1 d\mu_P(q) e^{-\frac{N}{2\sigma^2}[2X(P-q)+Y(q^2-p^2)]},
\label{ap:frac}
\eea
where
\bea
X&=&\frac{1}{N}\sum_{k=1}^N a_k r_k,\\
Y&=&\frac{1}{N}\sum_{k=1}^N a_k.
\eea
Conditionally on $P$, random variables are $X$ and $Y$ are equal to sums of independent identically distributed random variables and
\bea
\mathbb{E}(X\mid P=p)=\frac{p}{2},\\
\mathbb{E}(Y\mid P=p)=\frac{1}{2}.
\eea
Therefore, conditionally on $P$, $X$ and $Y$ converge strongly to their respective expectation values.
This motivates the introduction of new random variables strongly converging to zero:
\bea
X_N=X-\frac{p}{2},~Y_N=Y-\frac{1}{2}.
\eea
Using $X_N, Y_N$,
the problem of calculating  $\lim_{N\rightarrow \infty} \frac{1}{N}I(P;(A,R))$ can be formulated as follows: compute
\bea
L=-\lim_{N\rightarrow \infty} \frac{1}{N} \mathbb{E}_{P,X_N,Y_N} log F(P,X_N, Y_N),
\eea
where
\[
F(P,X,Y)=\int_{0}^1 d\mu_P(q)e^{-\frac{(p-q)^2}{4\sigma^2}-\frac{N}{2\sigma^2}(2X(p-q)+Y(q^2-p^2))}.
\]
Let us fix $\epsilon>0$.
A calculation which essentially repeats the derivation of Chebyshev inequality shows that
\bea
\Pr(X_N^2+Y_N^2\geq \epsilon \mid P=p)\leq \frac{\sigma^2+1}{2\epsilon^2 N}.
\label{ap:cheb}
\eea
Now let us re-write $L$ using the partition of unity as the sum of two terms:
\bea
L=\lim_{N\rightarrow \infty} (A_N+B_N),
\label{app:apb}
\eea
where
\bea
A_N=- \frac{1}{N} \mathbb{E}_{P,X_N,Y_N} \mathbf{1}_{X_N^2+Y_N^2\leq \epsilon}log F(P,X_N, Y_N),
\eea
\bea
B_N=- \frac{1}{N} \mathbb{E}_{P,X_N,Y_N} \mathbf{1}_{X_N^2+Y_N^2 > \epsilon}log F(P,X_N, Y_N),
\eea
Under the condition that the differential entropy (the logarithmic moment) of $p(J)$ exists, it is easy to show that
\bea
\mid A_N \mid \leq \frac{3}{2}\sigma^2 \epsilon.
\label{app:a}
\eea
Similarly,
\bea
\mid B_N\mid \leq \frac{1}{4\sigma^2}\mathbb{E}_{(X_N,Y_N,P)} (2|X_N|+|Y_N|)\mathbf{1}_{X_N^2+Y_N^2> \epsilon}\nonumber\\
\leq  \frac{1}{4\sigma^2} \sqrt{\mathbb{E}_{(X_N,Y_N,P)} (2|X_N|+|Y_N|)^2} \sqrt{Pr(X_N^2+Y_N^2> \epsilon)},\nonumber
\eea
where the last line is obtained using Schwarz inequality.
Using (\ref{ap:cheb}) and the fact that
$$
\mathbb{E}(2|X_N|+|Y_N|)^2=\frac{1}{N}\left( \frac{1}{2}+\sqrt{p^2+2\sigma^2}  \right)^2
$$
we find that
\bea
|B_N|\leq \frac{\sqrt{\mathbb{E}_P\left( \frac{1}{2}+\sqrt{p^2+2\sigma^2}  \right)^2}}{N\epsilon}.
\label{ap:b}
\eea
The expectation value in the above expression always exists as the random variable $P$ is bounded.
Combining (\ref{app:a}) and (\ref{ap:b}) we find that for any $\epsilon >0$ and any $N$,
$$
|A_N+B_N| \leq \frac{3}{2}\sigma^2 \epsilon+\frac{C}{\epsilon N}
$$
for some $\epsilon, N$- independent constant $C$.
Choosing $\epsilon=N^{-1/2}$ and taking the limit $N\rightarrow \infty$, we find that
$$
L=\lim_{N\rightarrow \infty} (A_N+B_N)=0.
$$

\section*{Acknowledgment}
The authors would like to thank Pr. David Wright (University of Exeter) and Dr. Charalampos Pozidis
(IBM Research, Zurich) for numerous useful discussions. Research has been partially supported
by the European Commission via FP$6$ collaborative grant ProTeM $2005-IST-5-34719$.

\ifCLASSOPTIONcaptionsoff
  \newpage
\fi

\end{document}